\documentclass[%
preprint,
superscriptaddress,
amsmath,
amssymb,
onecolumn,
longbibliography
]{revtex4-2}

\usepackage{amsmath}
\usepackage{amsfonts}
\usepackage{amssymb}
\usepackage{array}
\usepackage{bbm}
\usepackage{bm}
\usepackage{bbold}
\usepackage{braket}
\usepackage{bigints}
\usepackage{dsfont}
\usepackage[dvipsnames]{xcolor}
\usepackage{epsfig}
\usepackage{epstopdf}
\usepackage{pdfpages}
\usepackage{pgffor}
\usepackage{graphicx}
\usepackage{multirow}
\usepackage{mathtools}
\usepackage[T1]{fontenc}
\usepackage{times} 
\usepackage{tensor}
\usepackage[utf8]{inputenc}
\usepackage[unicode=true,breaklinks=true,colorlinks=true]{hyperref}

\hypersetup{citecolor=blue,urlcolor=blue}

\makeatletter 
\AtBeginDocument{\let\LS@rot\@undefined} 
\makeatother

\usepackage[normalem]{ulem}

\begin{document}

 


\title{Quantum delocalization on correlation landscape: The key to exponentially fast multipartite entanglement generation}

\author{Yaoming Chu}
\affiliation{School of Physics, Hubei Key Laboratory of Gravitation and Quantum Physics, International Joint Laboratory on Quantum Sensing and Quantum Metrology, Institute for Quantum Science and Engineering, Huazhong University of Science and Technology, Wuhan 430074, China}

\author{Xiangbei Li}
\affiliation{School of Physics, Hubei Key Laboratory of Gravitation and Quantum Physics, International Joint Laboratory on Quantum Sensing and Quantum Metrology, Institute for Quantum Science and Engineering, Huazhong University of Science and Technology, Wuhan 430074, China}

\author{Jianming Cai}
\email{jianmingcai@hust.edu.cn}
\affiliation{School of Physics, Hubei Key Laboratory of Gravitation and Quantum Physics, International Joint Laboratory on Quantum Sensing and Quantum Metrology, Institute for Quantum Science and Engineering, Huazhong University of Science and Technology, Wuhan 430074, China}

\begin{abstract}
Entanglement, a hallmark of quantum mechanics, is a vital resource for quantum technologies. Generating highly entangled multipartite states is a key goal in current quantum experiments. We unveil a novel framework for understanding entanglement generation dynamics in Hamiltonian systems by quantum delocalization of an effective operator wavefunction on a correlation landscape. Our framework establishes a profound connection between the exponentially fast generation of multipartite entanglement, witnessed by the quantum Fisher information, and the linearly increasing asymptotics of hopping amplitudes governing the delocalization dynamics in Krylov space. We illustrate this connection using the paradigmatic Lipkin-Meshkov-Glick model and highlight potential signatures in chaotic Feingold-Peres tops. Our results provide a transformative tool for understanding and harnessing rapid entanglement production in complex quantum systems, providing a pathway for quantum enhanced technologies by large-scale entanglement.
\end{abstract}

\maketitle

\newpage

\noindent
{\bf Introduction}\\
Entanglement, a distinctive property that sets the quantum realm apart from its classical counterpart, is a vital resource for the development of quantum technologies. One of the most paramount objectives in modern quantum experiments is the rapid generation of multipartite entanglement from readily available non-entangled states, particularly at an exponentially fast speed \cite{Lu2019,Eldredge2017}. This is crucial for two primary reasons. Firstly, multipartite entanglement is essential for unveiling foundational problems in quantum mechanics, such as the puzzling quantum-to-classical boundary \cite{Arndt2014}, quantum nonlocality \cite{Ardehali1992}, and a wide range of quantum many-body effects \cite{Amico2008}. The exponentially fast generation of multipartite entanglement with an ultra-short evolution time, which could enhance robustness against decoherence \cite{Zhang2024}, is of particular interest for exploring the fundamental question:  what is the maximum achievable macroscopic entanglement in realistic quantum systems \cite{Frowis2018}? Secondly, multipartite entanglement holds immense potential as a critical resource for revolutionizing modern information technologies. Harnessing large-scale entangled states underscores the power of groundbreaking advances over classical counterparts in the latest developments of quantum-enhanced precision measurement \cite{Pezze2018,Marciniak2022,Pedrozo2020,Greve2022,Bao2020}, high-speed computation \cite{Briegel2009,Song2019,Omran2019}, and secure communication \cite{Hillery1999,Zhao2004,Wehner2018}. The speed of entanglement generation is crucial for improving the duty ratio of quantum information processing tasks and sustaining quantum advantages  \cite{Tran2021PRX,Shi2024}, with approaching the elusive Heisenberg limit in quantum metrology \cite{Chu2023} as a notable example.

Unfortunately, not only the generation of entangled state but also the analysis of entangling dynamics in large-scale quantum experiments face unprecedented complexity and remain challenges. While the celebrated Lieb-Robinson bounds suggest the tantalizing possibility of exponentially fast entanglement generation with long-range interactions \cite{Tran2021,Guo2020}, they fall short of pinpointing explicit interacting many-body systems that can realize this ambitious objective. To tackle this pivotal problem, we propose an approach that synergistically integrates two foundational concepts in quantum information theory and quantum physics: quantum Fisher information (QFI) \cite{Braunstein1994,Toth2014} as a potent witness of multipartite entanglement \cite{Hyllus2012,Toth2012,Hauke2016,Guhne2009,Apellaniz2017}, and quantum delocalization that is a captivating phenomenon underpinning a plethora of cutting-edge quantum effects such as thermalization, scrambling, and hydrodynamics in many-body systems\cite{Swingle2018,Rigol2008,Keyserlingk2018,Rakovszky2018,Khemani2018}. By harnessing the power of QFI and quantum delocalization in tandem, we aim to elucidate the intricate dynamics of entanglement generation in complex quantum systems and identify specific systems that can achieve exponentially fast generation of multipartite entanglement.

In this work, we demonstrate that the intricate dynamics of QFI can be elegantly captured by delocalizing an effective 2D wavefunction on a meticulously crafted correlation landscape, defined within an efficient operator subspace. By exploiting the Krylov approach \cite{Parker2019} to construct this subspace, we establish a comprehensive paradigm for accessing the exponentially fast generation of multipartite entanglement. Our analysis uncovers an intriguing link between the exponentially fast entanglement generation and the linear increase of Lanczos coefficients in Krylov space, thereby providing a deeper understanding of the mechanisms driving rapid entanglement creation. We illustrate the power and versatility of our approach using the paradigmatic Lipkin-Meshkov-Glick model \cite{Bhattacharjee2022} and extend our analysis to generic chaotic dynamics, highlighting the broad applicability of our results in the context of exponentially fast entanglement production.
\vspace*{0.1cm}
\noindent
\textbf{General framework for characterizing QFI evolution in many-body systems}\\
\noindent
The quantum Fisher information (QFI) sets the ultimate limit for measurement precision in quantum metrology \cite{Braunstein1994}, quantifying the sensitivity of a quantum state $\rho$ to a unitary transformation generated by a Hermitian interrogation operator $\hat{\mathcal{O}}$ associated with a parameter $\vartheta$ to be estimated. The attainable parameter estimation uncertainty is constrained by the quantum Cramér-Rao bound, $(\Delta\vartheta)^2\geq 1/F_Q[\rho,\hat{\mathcal{O}}]$, where $F_Q[\rho,\hat{\mathcal{O}}]$ represents the QFI of the parametrized quantum state $\rho_{\vartheta}= e^{-i\vartheta \hat{\mathcal{O}}} \rho e^{i\vartheta \hat{\mathcal{O}}}$. To enhance sensing performance in a metrological task, an entangled input state $\rho$ exhibiting large QFI for $N>1$ particles is sought. Remarkably, the entanglement depth $\mathcal{E}_{\mathrm{d}}$ of a multipartite quantum state [namely the state is $\mathcal{E}_{\mathrm{d}}$-producible, but not $(\mathcal{E}_{\mathrm{d}}-1)$-producible] is delicately linked to the QFI density through $\mathcal{E}_{\mathrm{d}} \ge F_Q/N$ \cite{Hauke2016} for the standard scenario of a local $\hat{\mathcal{O}}=\sum_{i=1}^N \hat{o}_i$. Consequently, achieving the Heisenberg limit ($F_Q\simeq N^2$) necessitates preparing a globally entangled state with $\mathcal{E}_{\mathrm{d}} \sim N$. 

\begin{figure}[!t]
\centering
\includegraphics[width=13cm]{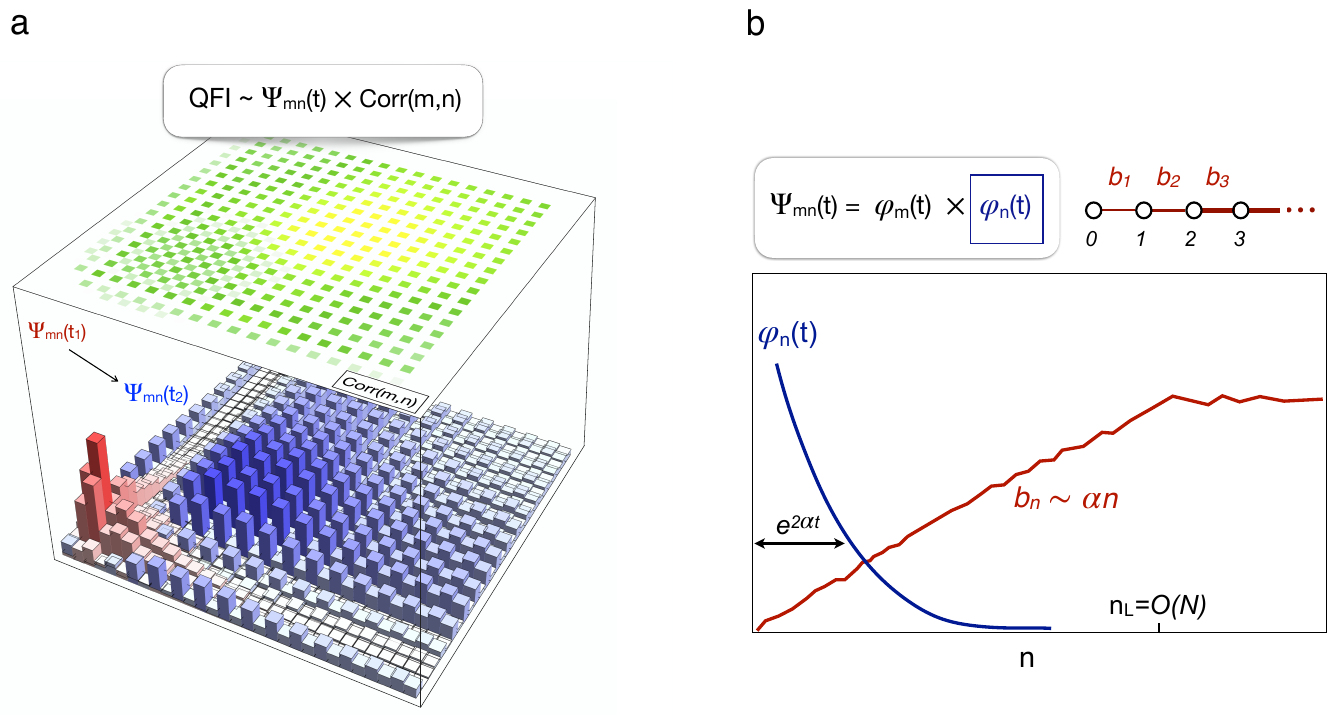}
\caption{\textbf{QFI evolution and exponentially fast entanglement generation from quantum delocalization picture.} \textbf{(a)} The QFI evolution can be interpreted as delocalizing a 2D wavefunction, $\Psi_{mn}(t)=\varphi_m(t) \varphi_n (t)$, on a correlation landscape denoted by $\mathrm{Corr}(m,n)$. It reaches the optimum when the main population of $\Psi_{mn}(t)$ evolves to the local maximum of $\mathrm{Corr}(m,n)$, i.e. marked by the light yellow region. \textbf{(b)} In Krylov space, the 1D component dynamics of $\Psi_{mn}(t)$ corresponds to quantum delocalization on a semi-infinite chain governed by a set of hopping amplitudes $\{b_n\}$. For many-body systems with a linearly increasing hopping amplitudes (i.e. $b_n\sim \alpha n$), a compelling exponentially fast delocalization of $\varphi_n(t)$ occurs with a spreading length of $\sim e^{2\alpha t}$. This characteristics is delicately connected to entanglement generation in an exponentially fast speed.}
\label{fig:principle}
\end{figure}

Here, we consider the preparation of an entangled state from a readily accessible product state $\sigma$ through unitary evolution governed by the Hamiltonian $\mathcal{H}$. The QFI of $\rho (t)= e^{-i \mathcal{H}t} \sigma  e^{i \mathcal{H} t}$, associated with the interrogation operator $\hat{\mathcal{O}}$, can be formulated in the following form \cite{Chu2023},
\begin{equation}
\label{Eq:QFI}
F_Q[\rho(t),\hat{\mathcal{O}}]=2 \gamma \mathrm{tr} ([\hat{\mathcal{O}}(t), \sqrt{\sigma}]^\dagger [\hat{\mathcal{O}}(t), \sqrt{\sigma}]),
\end{equation}
where $\gamma\in [1,2]$, and $\hat{\mathcal{O}}(t)=e^{i \mathcal{H}t} \hat{\mathcal{O}}  e^{-i \mathcal{H} t}$. This formula relates the QFI dynamics during the state preparation to the evolution of the interrogation operator in Heisenberg picture. Exploiting the Liouvillian superoperator, the time-evolved interrogation operator can be expresses as 
\begin{equation}
\mathcal{L}=[\mathcal{H},\bullet],\quad \hat{\mathcal{O}}(t) = e^{i\mathcal{L} t}\hat{\mathcal{O}}=\sum_{n=0}^{\infty} \frac{(i  t)^n}{n!} \mathcal{L}^n \hat{\mathcal{O}},
\end{equation}
where the second equality results from Baker-Campbell-Hausdorff formula. This expression indicates that the operator evolution is confined in an operator subspace spanned by the recursively generated commutators $\{\hat{\mathcal{O}},\mathcal{L}\hat{\mathcal{O}}, \mathcal{L}^2\hat{\mathcal{O}},\cdots\}$. Supposing that the Hermitian operator set $\{\hat{\mathcal{O}}_0,\hat{\mathcal{O}}_1,\hat{\mathcal{O}}_2,\cdots\}$ forms an orthogonal basis of such a subspace, we are able to expand the time-evolved operator as
\begin{equation}
\hat{\mathcal{O}}(t)=\sum_{n=0} \varphi_n(t) \hat{\mathcal{O}}_n,
\end{equation}
where the real coefficient vector, $[\varphi_0(t),\varphi_1(t),\varphi_2(t),\cdots]^T$, represents an effective operator wavefunction that obeys a Schrödinger-type equation governed by the Liouvillian \cite{supplement}. Its dynamics characterizes the delocalization of $\hat{\mathcal{O}}(t)$ on a one-dimensional chain formed by $\{\hat{\mathcal{O}}_n\}$. This basis can usually be interpreted as stratifying operators according to their ``complexity'', where $\mathcal{L}^n \hat{\mathcal{O}}$ becomes increasingly complex through recursive commutations with a many-body Hamiltonian. Physically, this process describes the generic delocalization of simple operators into an infinite ``bath'' of increasingly nonlocal operators, a phenomenon known as operator growth \cite{Parker2019,Nahum2018,Schuster2023}. 
As a major result of this work, we rigorously reformulate the general QFI evolution within this operator subspace as follows \cite{supplement}
\begin{equation}
\label{Eq:QFI_Ref}
F_Q[\rho(t),\hat{\mathcal{O}}]=4\sum_{m,n=0} \varphi_m(t) \varphi_n (t) \mathrm{Corr}(m,n),
\end{equation}
where $\mathrm{Corr}(m,n)$ represents a correlation function between the basis operators $\hat{\mathcal{O}}_m$ and $\hat{\mathcal{O}}_n$. Specifically, for an arbitrary pure initial state, it follows the form  \cite{supplement}
\begin{equation}
\mathrm{Corr}(m,n)=\langle \hat{\mathcal{O}}_m \hat{\mathcal{O}}_n+\hat{\mathcal{O}}_n \hat{\mathcal{O}}_m\rangle/2-\langle \hat{\mathcal{O}}_m \rangle \langle \hat{\mathcal{O}}_n\rangle,
\end{equation}
with the expectation $\langle \bullet \rangle$ taken with respect to the initial state. We define a 2D wavefunction, $\Psi_{mn}(t)=\varphi_m(t) \varphi_n (t)$, to characterize quantum delocalization along two orthogonal directions, each of which is independently quantified by the operator wavefunction. Thereby, the main result in Eq.\,\eqref{Eq:QFI_Ref} has a direct conceptual implication, unveiling an intriguing framework that the QFI dynamics can be generally interpreted as the delocalization of an effective wavefunction $\Psi_{mn}(t)$ on a correlation landscape characterized by $\mathrm{Corr}(m,n)$, as shown by Fig.\,\ref{fig:principle}(a).
\vspace*{0.3cm}
\noindent
\textbf{Guiding principle for identifying exponentially fast entanglement generation}\\
We primarily focus on the common scenario where the nested commutators $\{\hat{\mathcal{O}},\mathcal{L}\hat{\mathcal{O}}, \mathcal{L}^2\hat{\mathcal{O}},\cdots\}$ fail to close, indicating the involvement of an infinite-dimensional Lie algebra. This situation typically arises in genuine interacting many-body Hamiltonians. Explicitly, we employ the Lanczos algorithm \cite{Caputa2022}, which orthogonalizes $\{\mathcal{L}^n \hat{\mathcal{O}}\}$ through a Gram-Schmidt procedure. The algorithm yields a sequence of positive numbers, $\{b_n\}$, known as Lanczos coefficients, and an orthogonal sequence of Hermitian operators, $\{i^n\hat{\mathcal{O}}_n\}$, referred to as Krylov basis spanning the Krylov space. In this space, the Liouvillian takes the form of a tridiagonal matrix  \cite{supplement}:
\begin{equation}
\mathcal{L}=\left(
\begin{array}{ccccc}
0 & i b_1 & 0 & 0 & \cdots \\
-i b_1 & 0 & i b_2 & 0 & \cdots \\
0 & -i b_2 & 0 & i b_3 & \cdots\\
0 & 0 & -i b_3 & 0 & \ddots \\
\vdots & \vdots & \vdots & \ddots & \ddots\\
\end{array}\right).
\end{equation}
Straightforwardly, the operator wavefunction satisfies the following Schrödinger equation on a semi-infinite chain,
\begin{equation}
\label{Eq:WF_dynamics}
\partial_t \varphi_n =-b_{n+1}\varphi_{n+1}+b_n \varphi_{n-1}, \quad \varphi_{n}(0)=\delta_{n0},
\end{equation}
with $b_0=\varphi_{-1}=0$ by convention. Based on this equation, the universal properties of operator delocalization can be classified by different asymptotics of Lanczos coefficients  \cite{supplement}. A particularly intriguing case would be the linearly increasing one, namely $b_n\sim \alpha n$ for some real constant $\alpha >0$. This case leads to an exponentially fast delocalization of $\varphi_n(t)$ in Krylov space, which is well characterized by a quantity called  Krylov complexity to measure the expected position of $\varphi_n(t)$, namely $\mathcal{K}(t)=\sum_{n} n \varphi_n(t)^2 \sim e^{2\alpha t}$ \cite{Parker2019}. Moreover, such a characteristics can also be captured by an approximate form of $|\varphi_n(t)|\sim e^{-n/\xi(t)}$ at large $n$, where $\xi(t)\sim e^{2\alpha t}$ is a delocalization length that grows exponentially in time, as shown by Fig.\,\ref{fig:principle}(b).
Importantly, Eq.\,\eqref{Eq:QFI_Ref} reveals that the dynamical behavior of the QFI is determined by the delocalization of the operator wavefunction and the feature of the correlation landscape. 
We consider a paradigm in which the diagonal part of $\mathrm{Corr}(m,n)$ contributes dominantly to the QFI and exhibits a local maximum around a point $(n^*,n^*)$, as illustrated by the light yellow region in Fig.\,\ref{fig:principle}. This paradigm can be realized in various explicit scenarios and directly implies that the QFI reaches a local maximum when the main population of the operator wavefunction, $\varphi_n(t)$, arrives at the point $n=n^*$ in Krylov space. Based on the above analysis of operator delocalization, we can formulate the following proposition for identifying many-body interacting systems that enables exponentially fast entanglement generation.

\textbf{Proposition}: If the operator wavefunction $\varphi_n(t^*)$ at the evolution time $t=t^*$, when the QFI achieves its (local) maximum, is completely covered by the linearly increasing region of the Lanczos coefficients (denoted as $[0,n_{\mathrm{L}}]$), the optimal evolution time would satisfy $t^*\simeq \log [\mathcal{K}(t^*)] \lesssim \log n_{\mathrm{L}}\lesssim \log N$.

The last inequality in this proposition results from the fact that  the linear region of Lanczos coefficients is usually up to the order of $n_{\mathrm{L}}\lesssim  O(N)$ for a $N$-particle system \cite{Barbon2019}. We also point out a criteria that if $n_{\mathrm{L}}\gg n^*$, no significant population of $\varphi_n(t^*)$ would leak out of the region of $[0,n_{\mathrm{L}}]$, by noticing the exponentially decay tail of $\varphi_n(t)$ on the $n$ axis. This can be made more explicit, e.g. by requiring that $n_{\mathrm{L}}/ n^* \geq \nu$ with $\nu=4$  \cite{supplement}. We remark that such a proposition provides us a general guiding principle to find multipartite interacting systems that can create  entanglement in an exponentially fast speed by deterministic Hamiltonian evolution. Specifically,  we should engineer many-body Hamiltonians with  linearly increasing Lanczos coefficients, such as the ones describing scrambling and chaotic quantum systems, which covers previously studied explicit examples \cite{Kitagawa1993,Micheli2003,Kajtoch2015,Muessel2015,Munoz2023,Li2023,Zhang2024}. 

%

\begin{figure}[!t]
\centering
\includegraphics[width=15cm]{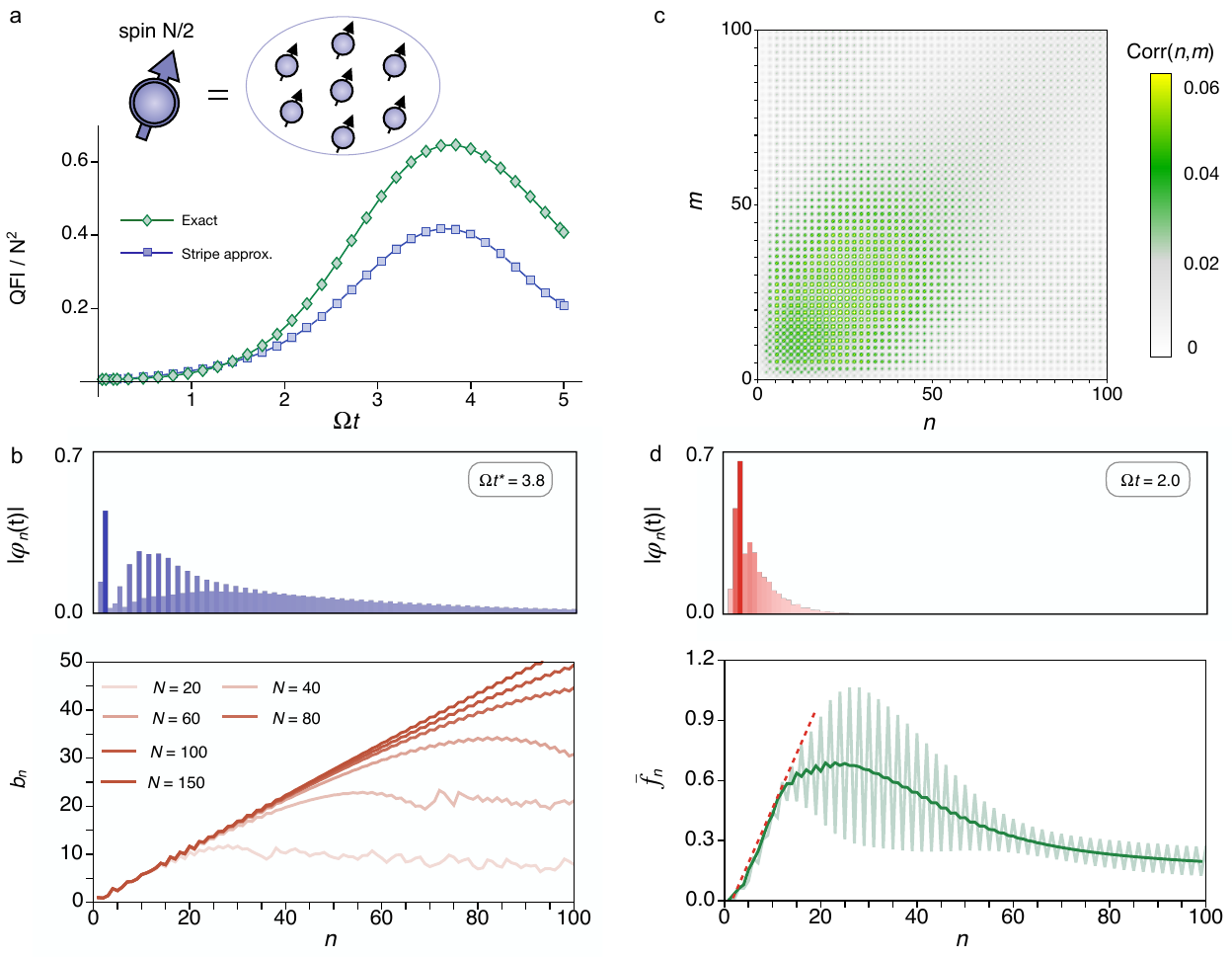}
\caption{\textbf{Exponentially fast generation of global entanglement in LMG model.} \textbf{(a)} The exact QFI and the QFI variant from the delocalization picture grow exponentially at transient times, and then slows down to achieve the optimum. \textbf{(b)} The Lanczos coefficients show  a linear increase up to $n_{\mathrm{L}} \simeq N$. At the time $t= t^*\approx 3.8/\Omega$ when the QFI is maximal, the main population of $\Psi_{mn}(t)$ reaches the dominant region of $\mathrm{Corr}(m,n)$ around $m^*=n^*\approx 25$  in panel \textbf{(c)}. The straightforward observation of $n_{\mathrm{L}}\geq 5 n^*$ leads to that $\varphi_n(t^*)$ is almost completely covered by $[0,n_{\mathrm{L}}]$, indicating that the maximum of the QFI is achieved exponentially fast. \textbf{(d)} The reduced quantity $\bar{f}_n$ from the dominant stripe of $\mathrm{Corr}(m,n)$ increases linearly for $n\leq n_c\approx 12$, resulting in a transient exponential blow-up of $\mathcal{F}_Q(t)$ as long as $\varphi_n(t)$ is fully located in the region of $[0,n_c]$, i.e. approximately before the time $t=2/\Omega$. We note that the values of $\mathrm{Corr}(m,n)$ are rescaled by a factor of $N^2$,  the system size is set as $N=150$, and the stripe width is chosen as $w=10$.}
\label{fig:LMG}
\end{figure}

%
\vspace*{0.3cm}
\noindent
\textbf{Entangling dynamics in Lipkin-Meshkov-Glick model}\\
As an illustrative example, we consider the Lipkin-Meshkov-Glick (LMG) Hamiltonian \cite{Bhattacharjee2022},
\begin{equation}
\mathcal{H}_{\mathrm{LMG}}=-\frac{\chi}{N}J_x^2-\Omega J_z,
\end{equation}
where $\boldsymbol{J}=(J_x,J_y,J_z)$ represents the total spin of the system comprised of $N$ spin-$1/2$ particles. Starting from a coherent spin state pointing along the positive $z$-axis (i.e $|\Psi\rangle=|\uparrow\uparrow\cdots\uparrow\rangle$), the system exhibits a saddle-point dominated scrambling dynamics at $\chi=2\Omega$. The QFI blows up exponentially at initial times, and subsequently slows down to achieve the maximum, $F_Q[\rho(t), J_x]\approx 0.64 N^2$, for an optimal evolution time $t= t^*$, see Fig.\,\ref{fig:LMG}\,(a). It can be seen that the operator wavefunction at the time $t^*$ is almost fully located in the linearly increasing region of Lanczos coefficients [Fig.\,\ref{fig:LMG}\,(b)], indicating that a globally entangled state of $F_Q/N\sim O(N)$ is generated in an exponentially fast speed, namely $t^*\lesssim \log N$. We confirm this result by exact numerical fitting in Supplementary Materials  \cite{supplement}. 


Furthermore, we demonstrate that the intricate growth behavior of the QFI can be understood from the picture of quantum delocalization in correlation landscape. As shown by the explicit calculation of $\mathrm{Corr}(m,n)$ in Fig.\,\ref{fig:LMG}\,(c), the diagonal stripe of the correlation landscape contributes dominantly to the QFI. We introduce the following quantity to characterize the feature of the correlation landscape,
\begin{equation}
f_n  =2\sum_{m=n-w}^{n-1}\mathrm{Corr}(m,n) +\mathrm{Corr}(n,n),
\label{Eq:scafn}
\end{equation}
where the integer $w$ is properly chosen according to the stripe width. One can see in Fig.\,\ref{fig:LMG}\,(d
) that $f_n$, depicted by the light green line, shows a linear increase up to a certain value of $n=n_c$. We take an average of $f_n$ between adjacent sites to eliminate the strong fluctuations for large $n$, and denote it as 
$\bar{f}_n=(f_n+f_{n+1})/2$, which achieves a maximum at $n\approx n^*$. By further making a replacement of $\varphi_n(t)\varphi_m(t)\to \varphi_n^2(t)$ in Eq.\,\eqref{Eq:QFI_Ref} over the diagonal stripe, we can define a variant of the QFI to qualitatively capture its dynamical behavior from a 1D delocalization picture, 
\begin{equation}
\mathcal{F}_Q =4\sum_{n=0}^{\infty}\bar{f}_n \varphi_n^2(t).
\label{Eq:scaF}
\end{equation}
This formulation predicts that $\mathcal{F}_Q$ is proportional to Krylov complexity and exhibits an exponential blow-up at initial times once $\varphi_n(t)$ is located in the region of $[0,n_c]$, namely $\sum_{n>n_c}^\infty \varphi_n(t)^2$ is negligible  [see Fig.\,\ref{fig:LMG}\,(d)]. Beyond this region, the growth of $\mathcal{F}_Q$ slows down and reaches its maximum when the main population of $\varphi_n(t)$ arrives the maximum of $\bar{f}_n$.

\begin{figure}[!t]
\centering
\includegraphics[width=15cm]{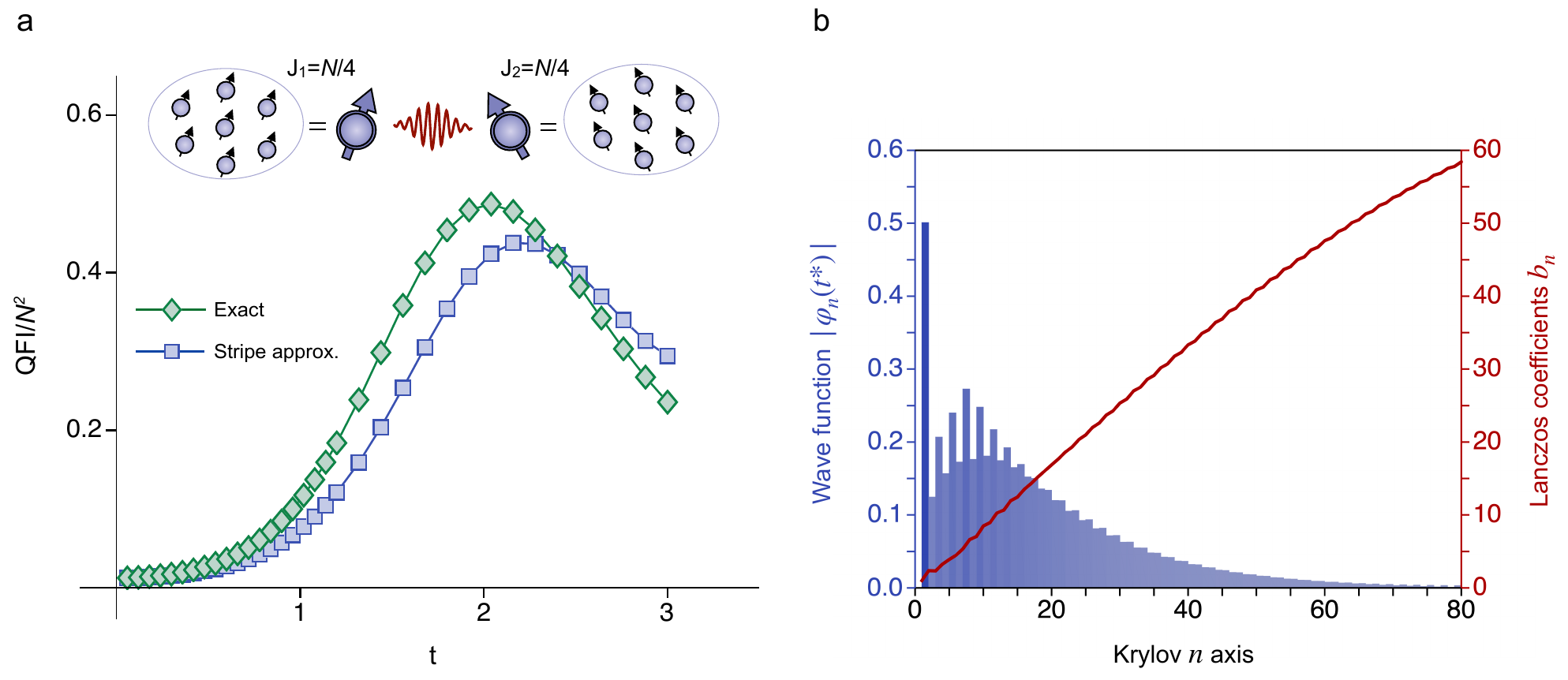}
\caption{\textbf{Exponentially fast generation of global entanglement in FP model.} \textbf{(a)} The exact QFI and its variant from 1D delocalization picture as defined in Eq.\,\eqref{Eq:scaF} exhibit similar behaviors, which grow exponentially at transient times and then slows down to achieve the optimum. \textbf{(b)} The operator wavefunction $\varphi_n(t^*)$ when the QFI achieves its maximum [$F_Q (t^*)\approx 0.48 N^2$ at $t^*\approx 2.03$] is fully covered by the linearly increasing region of Lanczos coefficients, indicating an exponentially fast generation of a globally entangled state. Here, the system size is $N=80$ and $J=N/4=20$, the stripe width is chosen as $w=30$.}
\label{fig:FP}
\end{figure}

\noindent
\textbf{Exponentially fast entanglement generation in chaotic FP model}\\
The present guiding principle points to versatile many-body systems with linearly increasing asymptotics of Lanczos coefficients for exponentially fast entanglement generation, with chaotic quantum systems as the most straightforward examples. A universal operator growth hypothesis states that an infinite, non-integrable, many-body system possesses an asymptotically linear Lanczos coefficients for an operator $\hat{\mathcal{O}}$ that has zero overlap with any conserved quantity \cite{Parker2019}. As a well-studied model of quantum chaos, the Feingold-Peres (FP) model of coupled tops describes a system of two spin-$J$ components, 1 and 2, of which the Hamiltonian is given by
\begin{equation}
\mathcal{H}_{\mathrm{FP}}=(1+c) (J_1^z+J_2^z)+\frac{4}{J}(1-c)J_1^x J_2^x,
\end{equation}
where the parameter $c\in[-1,1]$ and the spin operators $J_i^\alpha$ satisfy the $SU(2)$ algebra $[J_i^\alpha, J_j^\beta]=i\delta_{ij}\varepsilon_{\alpha \beta\gamma}J_i^\gamma$. It is noninteracting at $c=\pm 1$ and displays generic chaotic dynamics in the intermediate region \cite{Fan2017}. Remarkably, this system has not been exploited to prepare multipartite entanglement. Here we find that, starting from the initial product state  $|\Psi\rangle=|\uparrow\rangle^{\otimes{N/2}}\otimes |\uparrow\rangle^{\otimes{N/2}}$, the Lanczos coefficients at $c=0$ increase linearly up to the order of the spin size [see Fig.\,\ref{fig:FP} (a)]. An exponentially fast growth of the QFI associated with the interrogation operator $\hat{\mathcal{O}}=J_1^x+J_2^x$ that has no overlap with the system Hamiltonian highlights applicability of the guiding principle [see Fig.\,\ref{fig:FP} (b)]. Similar to the analysis in LMG model, we find that the QFI achieves its optimum $F_Q\approx 0.48 N^2$ at $t^*\lesssim \log(N)$, which achieves Heisenberg limit and might be utilized to efficiently construct two-partite quantum enhanced sensors \cite{Kaubruegger2023}. This result is also verified by exact numerical fitting in Supplementary Materials  \cite{supplement}.

\noindent
{\bf Conclusion $\&$ outlook}\\
To summarize, we have developed a comprehensive framework for quantifying the growth of multipartite entanglement in many-body systems by connecting the evolution of the QFI to quantum delocalization in an operator subspace. Using the efficient Krylov approach, we have established a link between the exponentially fast generation of multipartite entanglement and the universal characteristics of the linearly increasing Lanczos coefficients in generic chaotic or scrambling Hamiltonian systems. We remark that the general entangling dynamics in Eq.\,(4) is deeply connected to initial states involved in the correlation landscape. After identifying a multipartite interacting system, the subsequent construction of delicate optimization procedures to determine the optimal form of the initial state  deserves intense efforts in future research.
It should also be pointed out that the local-operator-related QFI might be blind to non-local entanglement embedded in the celebrated W states and topological phases \cite{Hauke2016}. A remain challenge is the question whether non-local extensions of the presented framework permit the efficient analysis of entanglement growth speed in relevant strategies \cite{Bugu2013,Ozaydin2021}. 
Other interesting directions include extending our framework to periodically driven chaotic systems \cite{Reilly2023,Liu2021} as well as open quantum systems with decoherence \cite{supplement}, and exploring the potential of our results to identify signatures of rapidly generating macroscopic quantum states \cite{Frowis2018,Frowis2012}. The implications of our findings extend beyond the realm of fundamental quantum physics, as the ability to generate entanglement exponentially fast has far-reaching consequences for the development of cutting-edge quantum technologies. 

\noindent
{\bf Acknowledgements}\\
This work is supported by the National Natural Science Foundation of China (12161141011 and 12174138), the National Key R$\&$D Program of China (2018YFA0306600), and the Shanghai Key Laboratory of Magnetic Resonance (East China Normal University). Y.-M.C. is also supported by the Young Scientists Fund of the National Natural Science Foundation of China (12304572) and the fellowship of China Postdoctoral Science Foundation (2022M721256). 


%

\onecolumngrid
 

\section*{Supplementary material}

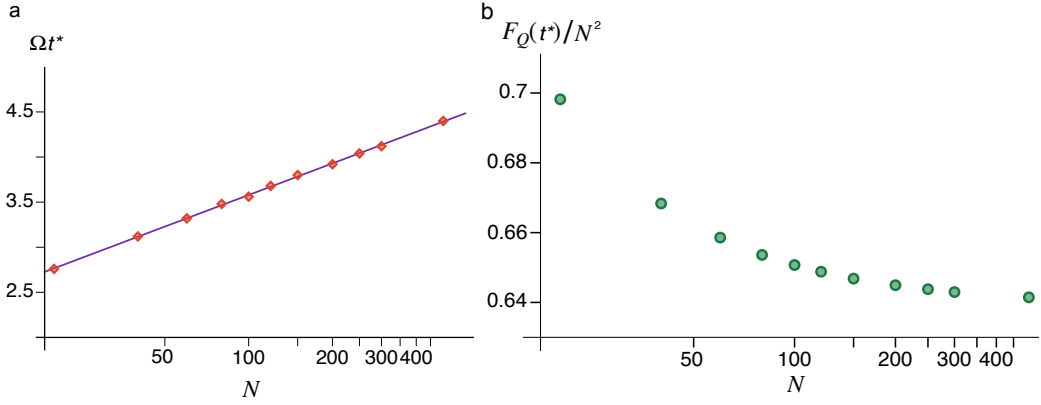

FIG. S1. **Numerical fitting of exponentially fast generation of global entanglement in LMG model.** (a) The evolution time $t^*$ at which the QFI achieves its maximum shows a logarithmic relation with the system size as $\Omega t^* \approx 0.51 \log N + 1.25$. (b) The maximal QFI at the time $t^*$ approximately saturates to $F_Q(t^*) \approx 0.64 N^2$ in the large-$N$ limit. The system parameter is set as $\chi = 2\Omega$.

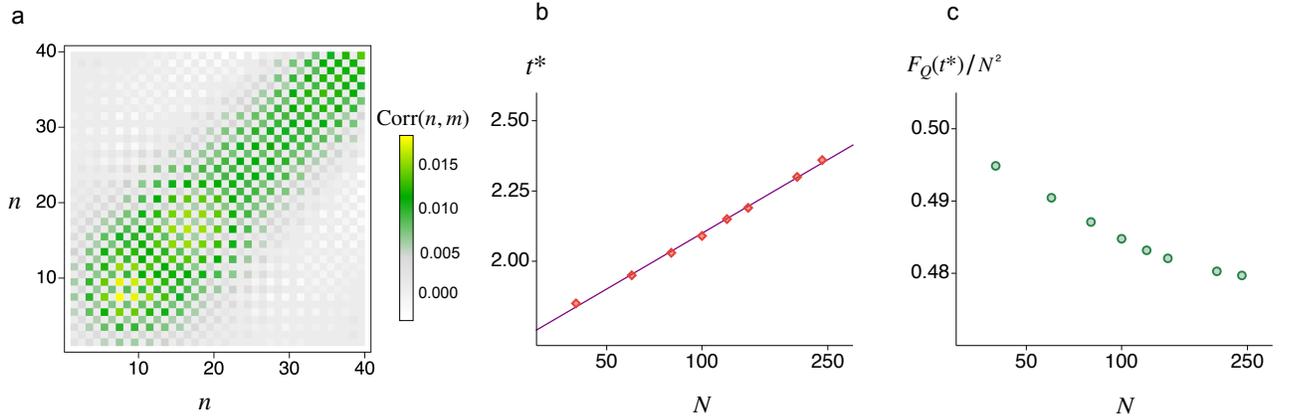

FIG. S2. **Extended materials for Feingold-Peres model.** (a) The correlation landscape is dominated by its diagonal stripe and exhibit a local maximum around $(n^*, n^*) \approx (10, 10)$ for the system size $N = 80$. Combined with Fig. 3 in the main text, one can find that $n_L \gtrsim 4n^*$ with $n_L$ characterizing the linearly increased region of the Lanczos coefficients. This implies an exponentially fast generation of global multipartite entanglement. (b) The evolution time $t^*$ at which the QFI achieves its optimum shows a relation with the system size as $t^* \approx 0.29 \log N + 0.78$. (c) The maximal QFI at the time $t^*$ approximately saturates to $F_Q(t^*) \approx 0.48 N^2$ in the large-$N$ limit. The system parameter of FP model is chosen as $c = 0$.

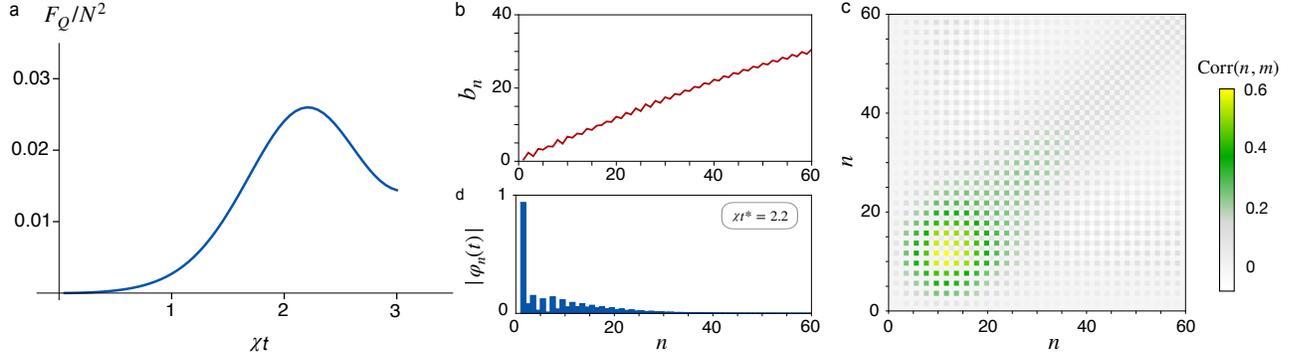

FIG. S3. **Exponentially fast entanglement generation by $SU(1,1)$ approach.** (a) The QFI achieves its first local maximum around $\chi t^* = 2.2$. (b) The Lanczos coefficients $\{b_n\}$ show a linear increase up to the order of $n_\mathrm{L} \simeq N$. (c) The correlation landscape $\mathrm{Corr}(n,m)$ is dominated by its diagonal stripe and exhibit a local maximum around $(n^*, n^*) \approx (15, 15)$. (d) The fact that $n_\mathrm{L} \gtrsim 4n^*$ leads to that the operator wavefunction at $t^* \approx 2.2/\chi$ when the QFI is approximately optimized, is fully covered by the linearly increased region of the Lanczos coefficients, implying that the first local maximum of the QFI can be achieved exponentially fast. We choose the particle number as $N = 60$ and the system parameters as $\chi = 1$ and $q = -1$.

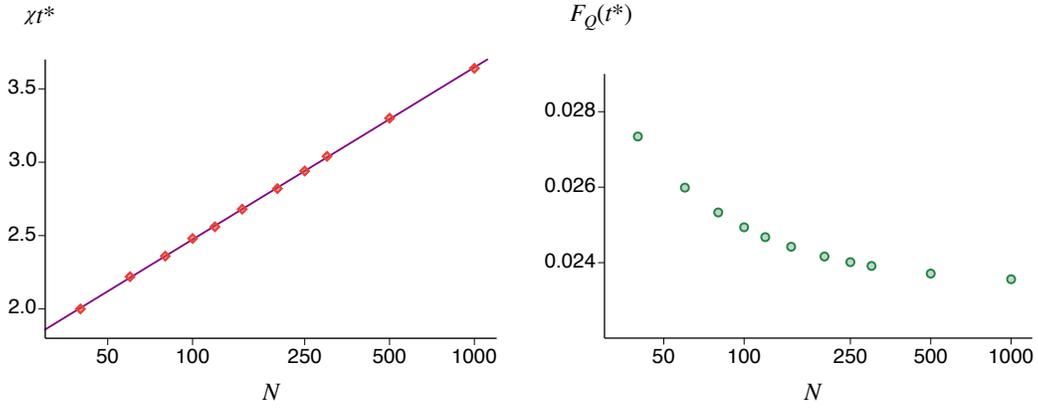

FIG. S4. **Numerical fitting of exponentially fast entanglement generation by $SU(1,1)$ approach.** (a) The evolution time $t^*$ at which the QFI achieves its optimum shows a logarithmic relation with the particle number of the system, i.e. $\chi t^* \approx 0.51 \log N + 0.13$. (b) The maximal QFI at the time $t^*$ approximately saturates to $F_Q(t^*) \approx 0.024 N^2$ in the large-$N$ limit. We set the system parameters as $\chi = 1$ and $q = -1$.



\end{document}